\documentclass[a4paper,11pt]{article}

\usepackage[utf8]{inputenc}
\usepackage[T1]{fontenc}

\usepackage{jcapmod}

\usepackage{amsmath,amsfonts,amssymb}
\usepackage{mathtools,physics,bm}
\usepackage{textalpha}

\usepackage{hyperref}
\usepackage{graphicx}

\usepackage[a4paper,margin=2.55cm]{geometry}

\numberwithin{equation}{section}

\def\DHLhksqrt#1#2{%
\setbox0=\hbox{$#1\sqrt{#2\,}$}\dimen0=\ht0
\advance\dimen0-0.2\ht0
\setbox2=\hbox{\vrule height\ht0 depth -\dimen0}%
{\box0\lower0.4pt\box2}}

\makeatletter
\DeclareFontFamily{OMX}{MnSymbolE}{}
\DeclareSymbolFont{MnLargeSymbols}{OMX}{MnSymbolE}{m}{n}
\SetSymbolFont{MnLargeSymbols}{bold}{OMX}{MnSymbolE}{b}{n}
\DeclareFontShape{OMX}{MnSymbolE}{m}{n}{
    <-6>  MnSymbolE5
   <6-7>  MnSymbolE6
   <7-8>  MnSymbolE7
   <8-9>  MnSymbolE8
   <9-10> MnSymbolE9
  <10-12> MnSymbolE10
  <12->   MnSymbolE12
}{}
\DeclareFontShape{OMX}{MnSymbolE}{b}{n}{
    <-6>  MnSymbolE-Bold5
   <6-7>  MnSymbolE-Bold6
   <7-8>  MnSymbolE-Bold7
   <8-9>  MnSymbolE-Bold8
   <9-10> MnSymbolE-Bold9
  <10-12> MnSymbolE-Bold10
  <12->   MnSymbolE-Bold12
}{}

\let\llangle\@undefined
\let\rrangle\@undefined
\DeclareMathDelimiter{\llangle}{\mathopen}%
                     {MnLargeSymbols}{'164}{MnLargeSymbols}{'164}
\DeclareMathDelimiter{\rrangle}{\mathclose}%
                     {MnLargeSymbols}{'171}{MnLargeSymbols}{'171}
\makeatother

\begin{document}


\begin{titlepage}

\baselineskip=15.5pt \thispagestyle{empty}

\begin{center}
    {\fontsize{15}{24}\selectfont \bfseries Delicate curvature bounces \\ \vspace{5pt} in the no-boundary wave function and in the late universe}
\end{center}

\vspace{0.1cm}

\begin{center}
    {\fontsize{12}{18}\selectfont Jean-Luc Lehners$^{1}$ and Jerome Quintin$^{2,3}$}
\end{center}

\begin{center}
    \vskip8pt
    \textsl{$^1$ Max Planck Institute for Gravitational Physics (Albert Einstein Institute),\\
    Am M\"uhlenberg 1, D-14476 Potsdam, Germany}
    \vskip4pt
    \textsl{$^2$ Department of Applied Mathematics and Waterloo Centre for Astrophysics,\\
    University of Waterloo, 200 University Ave.~W., Waterloo, Ontario N2L 3G1, Canada}\\
    \vskip4pt
    \textsl{$^3$ Perimeter Institute for Theoretical Physics,\\
    31 Caroline St.~N., Waterloo, Ontario N2L 2Y5, Canada}\\
\end{center}

\vspace{1.2cm}

\hrule
\vspace{0.3cm}
\noindent {\bf Abstract}\\[0.1cm]
Theoretical considerations motivate us to consider vacuum energy to be able to decay and to assume that the spatial geometry of the universe is closed. Combining both aspects leads to the possibility that the universe, or certain regions thereof, can collapse and subsequently undergo a curvature bounce. This may have occurred in the very early universe, in a pre-inflationary phase. We discuss the construction of the corresponding no-boundary instantons and show that they indeed reproduce a bouncing history of the universe, interestingly with a small and potentially observable departure from classicality during the contracting phase. Such an early bouncing history receives a large weighting and provides competition for a more standard inflationary branch of the wave function. Curvature bounces may also occur in the future. We discuss the conditions under which they may take place, allowing for density fluctuations in the matter distribution in the universe. Overall, we find that curvature bounces require a delicate combination of matter content and initial conditions to occur, though with significant consequences if these conditions are met.  
\vskip10pt
\hrule
\vskip10pt

\end{titlepage}


\tableofcontents
\pagenumbering{arabic}
\setcounter{page}{1}


\section{Introduction}

A generic prediction of the no-boundary proposal, which is currently the best understood proposal for the quantum creation of the universe, is that the universe should be closed and have positive spatial curvature \cite{Hawking:1981gb,Hartle:1983ai} (for a review see \cite{Lehners:2023yrj}). This is because positive spatial curvature is required to close the geometry off in a smooth manner. This prediction can be in agreement with current upper bounds on the magnitude of the spatial curvature if the universe expanded sufficiently, so as to dilute the curvature appropriately. 

This last requirement is (at least at first sight) at odds with the other generic prediction of the no-boundary proposal, which is that an early inflationary phase is predicted to last a very short time (for a recent discussion see \cite{Maldacena:2024uhs}). This is because the no-boundary amplitude scales roughly as $|\Psi| \sim \exp[12 \pi^2/(\hbar V_\mathrm{ini})]$, where $V_\mathrm{ini}$ is the initial value of the scalar potential. Being low on the potential is then favoured. 

In the past, an anthropic argument was put forward to resolve this problem \cite{Hawking:2006ur,Hartle:2009ig}. However, it remains unclear how precise this argument can be made. Recently, two new (a priori not mutually exclusive) proposals were discussed for how one might remedy the problem of nulceating low on the potential. The first comes from the realisation that in a gravitational path integral, it might not make sense to sum over all metrics, as many metrics would lead to a divergence of the path integral once matter terms are included. Rather, the idea (first put forward by Kontsevich \& Segal in the context of quantum field theory defined on curved spacetime \cite{Kontsevich:2021dmb}) is to only allow complex metrics on which generic $p$-form matter fields are well defined. It turns out that this criterion requires the scalar potential to be very flat since in such a case the no-boundary instanton is `almost' real valued \cite{Lehners:2022xds,Hertog:2023vot}. 
Requiring $60$ $e$-folds of inflation and allowable metrics, it was found in \cite{Hertog:2023vot} that this requires the potential to be very flat, which translates into an upper bound on the tensor-to-scalar ratio. Turning the argument around, for several classes of phenomenologically realistic (e.g., plateau-type) potentials, one can show that the scalar field is required to nucleate above a certain threshold value, which happens to lie around the $60$ $e$-fold mark \cite{Lehners:2023pcn}.
Moreover, the inflationary phase would then not be much longer than required in order to solve the flatness problem, implying that the spatial curvature of the universe might be close to the upper bound stemming from observations \cite{Lehners:2023pcn}.

\begin{figure}[t]
	\centering
	\includegraphics[width=0.6\textwidth]{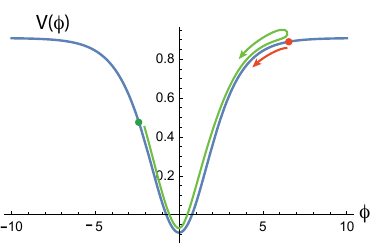}
	\caption{{\footnotesize A scalar potential $V(\phi)$ allowing two distinct early histories of the universe: in red, a standard inflationary history, and in green an alternative history in which the universe nucleates lower on the potential, then undergoes a recollapse when the potential turns negative, followed by a curvature bounce. The scalar field then eventually reaches the inflationary plateau on the right hand side, and from then on the two histories proceed identically.}}
	\label{fig:pot}
\end{figure}

\begin{figure}[t]
	\centering
    \includegraphics[width=0.6\textwidth]{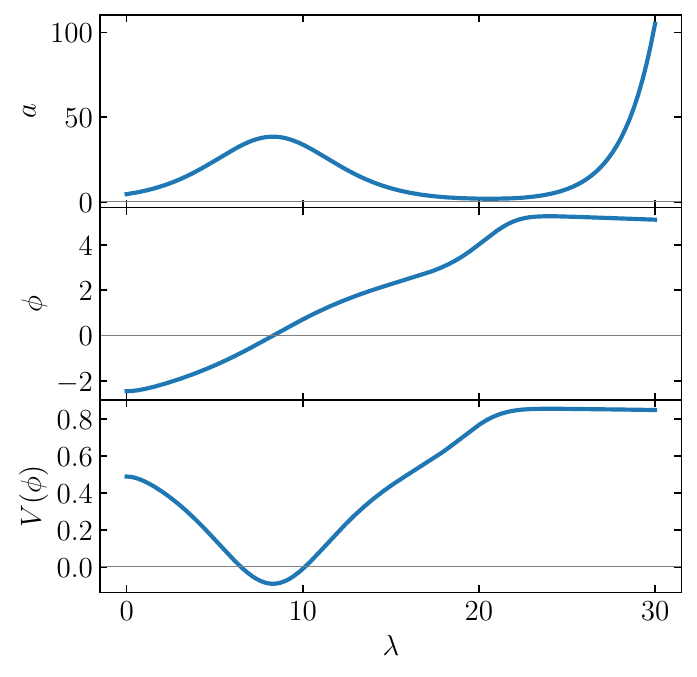}
	\caption{{\footnotesize A plot of the bouncing history. The top panel shows the evolution of the scale factor $a$ as a function of the time $\lambda$, the middle panel that of the scalar field $\phi$ and at the bottom we can see the evolution of the potential $V(\phi)$. This provides an alternative to a more standard purely inflationary history, which would directly start at the point where the scalar $\phi$ turns around at $\lambda \approx 24$. From there on the two histories become identical. The potential that we used is given in Eq.~\eqref{eq:pot}, and the initial conditions for the bouncing history are $a(0)=10/\sqrt{5}$, $\phi(0) = -\sqrt{6}$ and $\dot\phi(0)=0.$}}
	\label{fig:class}
\end{figure}

\begin{figure}[t]
	\centering
	\includegraphics[width=0.6\textwidth]{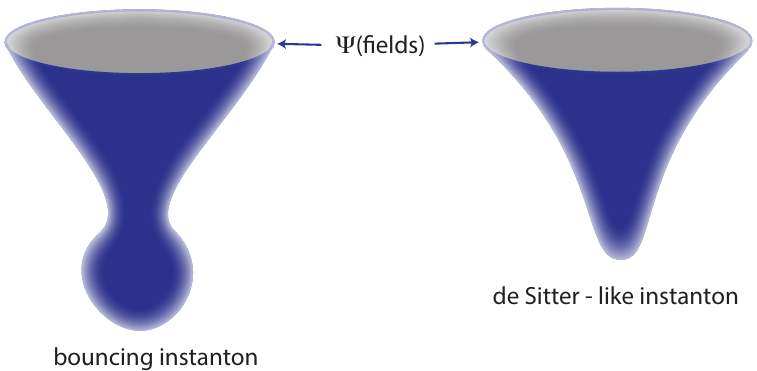}
	\caption{{\footnotesize A cartoon of the two types of instantons that we will consider: the ordinary inflationary instantons are sketched on the right, while the left graph shows the new type of `bouncing' instanton, in which the universe undergoes a recollapse and a curvature bounce at early times.}}
	\label{fig:cartoon}
\end{figure}

A second scenario was proposed by Matsui et al.~\cite{Matsui:2019ygj,Matsui:2023wxm,Matsui:2023ezh}. Their idea is that the universe could indeed nucleate fairly low on the potential, but that the universe might recollapse and bounce due to the positive spatial curvature. During such a curvature bounce, the scalar field can roll significantly up the potential and end up at a potential value that is higher than that at nucleation, where a proper inflationary phase could then take place. See Fig.~\ref{fig:pot} for an illustration of the potential, and Fig.~\ref{fig:class} for plots of the intended classical field evolution. This scenario thus posits a pre-history to inflation. In these works, no explicit construction of the corresponding no-boundary instantons and of the wave function was given. A first goal of the present paper is to do precisely this. Instantons of this type have a similar shape to the ususal inflationary no-boundary instantons, except that they have a bulge and a waist right after the nucleation phase of the universe --- see Fig.~\ref{fig:cartoon} for an illustration. Finding such instantons numerically presents new challenges, and we will describe their resolution below. 

Equipped with an explicit calculation of the wave function, we can address several questions. The first one is whether such `bouncing' instantons in fact lead to the prediction of a classical bouncing spacetime. This is non-trivial, as the prediction of classical spacetime is by no means automatic in quantum cosmology. And in fact, as we will discuss in some detail, the recollapse phase of the proposed history leads to a slight loss of classicality.
A second question of interest is whether the proposed bounce history comes out as preferred. Since we are dealing with a quantum theory of the early universe, the wave function contains other possible histories, in particular one that starts directly on the inflationary plateau --- see Fig.~\ref{fig:pot} for an illustration. We will contrast both types of history and discuss to what extent our current knowledge of quantum gravity allows us to assess their relative likelihood.

If the universe is created with a positive average spatial curvature, then it will retain that curvature, even though it becomes diluted by the expansion of the universe. Still, it is then conceivable that the curvature might play an important role again at a later stage of evolution. In this vein, we will analyse the conditions under which our universe might undergo a recollapse and bounce in the future. There are two important distinctions between the late and early bounces: at late times, the evolution of the universe is evidently classical, hence it is enough for us to analyse the classical equations of motion in this case. However, the presence of matter affects the conditions for bounces rather strongly. For this reason, we will extend our analysis to allow not just for matter, but also for both positive and negative density contrasts in addition, in order to assess which regions of the universe might be most likely to undergo a future bounce.

The plan of this paper is as follows: we will start with a brief overview of the salient features of the no-boundary proposal that we will require. In particular, this includes a discussion of the interpretation of the wave function and of the conditions for classicality. Then, in section \ref{sec:bi} we will present the construction of bouncing instantons and discuss their peculiarities. In section \ref{sec:ch}, we can then compare the implied bouncing history with a more usual purely inflationary phase. We will then turn our attention to curvature bounces at late times in section \ref{sec:late}, with a discussion of the conditions for recollapse and bounce in section \ref{sec:cb} and an extension to under- and overdense regions in section \ref{sec:density}. Our conclusions are presented in section \ref{sec:discussion}. We will use natural units where $c=1$, $8\pi G_\mathrm{N}=1, \hbar=1$, but occasionally retain $\hbar$ explicitly for clarity.

\section{Early bounces}

\subsection{The no-boundary framework} \label{sec:nb}

It seems likely that in order to understand the early stages of the universe we must treat the entire universe as a quantum system. And even though a full theory of quantum gravity remains unavailable at present, we may (tentatively) combine general relativity with quantum principles --- to the extent that we can --- to go beyond the classical picture of a singular big bang. As long as the spacetime curvature remains sufficiently small, one may expect this approach to provide a trustworthy approximation. 

An influential proposal in this vein is the Hartle-Hawking no-boundary proposal, which uses the path integral approach to quantise gravity (semi-classically) \cite{Hawking:1981gb,Hartle:1983ai}. A general question for the path integral approach is what kind of metrics one should sum over. The suggestion made by Hartle and Hawking was to sum over compact metrics that contain no boundary other than the present-day hypersurface. If these metrics (and the associated matter configurations) are also regular, then they will yield a finite action contribution to the wave function \cite{Jonas:2021xkx}. If they are stationary points of the path integral, then they will contribute significantly to the path integral. And the stationary point with the highest weighting will dominate the wave function.\footnote{We should note that the original suggestion to define the path integral as a sum over Euclidean metrics does not work, as the integral diverges. Summing over Lorentzian metrics also does not work \cite{Feldbrugge:2017kzv,Feldbrugge:2017fcc,Feldbrugge:2017mbc}, while proposals that involve sums over classes of complex metrics do seem to yield the appropriate no-boundary wave function \cite{Halliwell:1988ik,Janssen:2019sex,DiTucci:2019dji,Lehners:2024kus,Anninos:2024iwf}. A general definition has however not been worked out at this time. All we will need here are the features of the saddle points themselves, and these are unaffected by the choice of integration contours.}

Thus we must first look for compact and regular solutions of the field equations. In general these solutions turn out to be complex valued, a point to which we will return repeatedly. We will work in a symmetry reduced setting, where the metric is taken to be of closed Robertson-Walker form,
\begin{equation}
    \dd s^2 = -N^2 \dd t^2 + a(t)^2 \dd \Omega_3^2\,,\label{eq:ds2k1}
\end{equation}
where $\dd \Omega_3^2$ is the metric on the unit $3$-sphere, $N$ is the lapse function and $a$ the scale factor. As matter content, we will simply consider a scalar field $\phi(t)$ with a potential $V(\phi)$. Then the action of general relativity with the scalar is given by
\begin{equation}
    S = 2\pi^2 \int \mathrm{d}t~N\left( -\frac{3}{N^2}a\dot{a}^2 + \frac{1}{2 N^2}a^3 \dot\phi^2 +3a - a^3 V(\phi)\right)\,, \label{scalarmodel}
\end{equation}
where dots denote time derivtives. This action is of the minisuperspace form
\begin{equation}
    S = \int \dd t~N \left( \frac{1}{2} G_{AB} \frac{1}{{N}}\frac{\mathrm{d}q^A}{\mathrm{d}t}\frac{1}{N}\frac{\mathrm{d}q^B}{\mathrm{d}t} - U(q^A) \right)\,,\label{minisupact}
\end{equation}
where $q^A=(a,\phi)$ are the dynamical fields, with field space metric $G_{aa} = -12\pi^2 a$, $G_{\phi\phi} =  2\pi^2 a^3$ and effective potential $U = 2\pi^2( -3a + a^3 V)$.
The Hamiltonian density associated with this action is
\begin{equation}
    {\cal H} = \frac{1}{2} G^{AB} p_A p_B + U \,, \label{Hamiltonian}
\end{equation}
with the canonical momenta
\begin{equation}
    p_a = -12\pi^2a \frac{\dot{a}}{N}\,, \, \quad p_\phi = 2\pi^2 a^3 \frac{\dot\phi}{N}\,.
\end{equation}
For more details regarding this procedure and what follows, see the review \cite{Lehners:2023yrj}.

The Hamiltonian is classically zero, and this condition is equivalent to the Friedmann equation. One can quantise the theory canonically (in the field representation) by the replacement
\begin{equation}
    p_A \rightarrow -i\hbar \frac{\partial}{\partial q^A} \equiv -i\hbar \partial_A\,,
\end{equation}
which results in the Wheeler-DeWitt (WdW) equation
\begin{equation}
    \hat{\cal H} \Psi = \left( - \frac{\hbar^2}{2} \Box + U \right) \Psi = 0\, , \label{WdW}
\end{equation}
where $\Psi = \Psi(a,\phi)$ is the wave function of the universe, and where the field space d'Alembertian can be taken as $\Box=G^{AB}\nabla_A\nabla_B$. Note that the wave function is a function of the fields only, not of the coordinates. Also, in the no-boundary proposal, it is a function only of the fields on the final hypersurface. We will denote the corresponding values by $b \equiv a(t_\mathrm{final})$, $\chi \equiv \phi(t_\mathrm{final})$. So in our case we will have $\Psi = \Psi(b,\chi)$ for a series of $(b,\chi)$ values. Also, one can show that the path integral automatically solves the WdW equation, but with the advantage that it allows for a transparent implementation of the boundary conditions \cite{Halliwell:1988wc}. 

Let us now briefly review how one may interpret the wave function, as this will be of relevance below. First recall that the WdW equation admits a conserved current, defined as
\begin{equation}
    J^A = -\frac{i\hbar}{2}\left(\Psi^\star \nabla^A \Psi - \Psi \nabla^A \Psi^\star \right)\,, \label{current}
\end{equation}
which is conserved ($\nabla_A J^A=0$) subject to the WdW equation. We can make progress by writing the wave function in a suggestive form,
\begin{equation}
    \Psi = \exp\left[\frac{1}{\hbar}\left({\cal W} + i{\cal S}\right)\right]\,, \label{wfsuggestive}
\end{equation}
where ${\cal W}$ and ${\cal S}$ are real functions of the fields $q^A$. The function ${\cal W}$ is called the weighting and ${\cal S}$ the phase. Now one can expand the WdW equation \eqref{WdW} as a series in $\hbar$, with the two leading orders (real and imaginary parts) resulting in the equations
\begin{subequations}\label{WdW12}
\begin{eqnarray}
    -\frac{1}{2} (\nabla {\cal W})^2+\frac{1}{2} (\nabla {\cal S})^2 + U = 0\,, && \quad \nabla {\cal W} \cdot \nabla {\cal S} = 0\,, \label{WdW1} \\ \Box {\cal W} =0\,, && \quad \Box {\cal S} = 0\,, \label{WdW2}
\end{eqnarray}
\end{subequations}
where the shorthand notation stands for $\nabla {\cal W}\cdot \nabla {\cal S}\equiv G^{AB} \nabla_A {\cal W} \nabla_B {\cal S}$, $(\nabla {\cal W})^2\equiv (\nabla {\cal W})\cdot(\nabla {\cal W})$, and so on. The conserved current becomes
\begin{equation}
    J^A =  e^{2{\cal W}/\hbar}\, \nabla^A {\cal S} \,. \label{current2}
\end{equation}

Now,  if ${\cal W}$ varies slowly compared to ${\cal S}$, then we obtain the Hamilton-Jacobi equation of classical mechanics \cite{Vilenkin:1988yd},
\begin{equation} \label{eq:WKBcond}
    (\nabla {\cal W})^2 \ll  (\nabla {\cal S})^2 \quad \implies \quad \frac{1}{2} (\nabla {\cal S})^2 + U \approx 0\,,
\end{equation}
with ${\cal S}$ being the classical action and with the canonical momentum 
\begin{equation}
    p_A = \frac{\partial {\cal S}}{\partial q^A}\,. \label{momassign}
\end{equation}
This is nothing else than the Wentzel-Kramers-Brillouin (WKB) semi-classical approximation. The remaining equations in \eqref{WdW12} then guarantee the conservation of the current up to sub-leading order.

There are three important implications for us. The first is that in the WKB regime a stationary (in fact saddle) point contributes a term that is well approximated by
\[
    \exp\left[\frac{1}{\hbar}\left({\cal W}_\textrm{o-s}+i{\cal S}_\textrm{o-s}\right)\right]
\]
to the wave function $\Psi$, where the subscript `$\textrm{o-s}$' indicates that the action should be evaluated on-shell, i.e., on the solution to the equations of motion. The second is that the current can be used to define a relative probability
\begin{equation}
     {\cal P} = e^{2{\cal W}/\hbar} \, n^A\nabla_{A} {\cal S} \approx e^{2{\cal W}/\hbar} = \Psi^\star \Psi\,,
\end{equation}
where the $n^A$ direction is taken to be orthogonal to constant-${\cal S}$ surfaces. The third is that the associated classical history is specified by the relation
\begin{equation}
    p_A \Psi = -i\hbar \, \partial_A \Psi \approx \partial_A {\cal S} \, \Psi \quad  \implies \quad p_A \approx \partial_A {\cal S}\,.\label{pfromS}
\end{equation}
We will use these relations to assess whether the wave function is indeed describing a classical spacetime or not. Note that this is by no means guaranteed. At the moment, only two dynamical mechanisms are known that can drive the wave function to WKB form in the early universe \cite{Lehners:2015sia}: one is inflation \cite{Hartle:2008ng} and the other one is ekpyrosis \cite{Battarra:2014kga}. However, both mechanisms need to operate for some time to be effective, typically at least over a few $e$-folds. This will be illustrated in the example below.

\subsection{Bouncing instantons} \label{sec:bi}

We will now specialise to the potential 
\begin{equation}
    V(\phi) = \alpha \tanh\left(\frac{\phi}{\sqrt{6}}\right)^2 + \beta \tanh\left(\frac{\phi}{\sqrt{6}}\right) + \gamma\,, \label{eq:pot}
\end{equation}
which is of the form chosen by Matsui et al.~\cite{Matsui:2019ygj,Matsui:2023ezh,Matsui:2023wxm}. These authors made the choice $\alpha=1$, $\beta=0$, $\gamma = -0.09143$, where the coefficients were chosen such that a classical solution featuring a recollapse and bounce phase is obtained (in \cite{Matsui:2023ezh,Matsui:2023wxm} dissipative effects were also modelled, which we will ignore here). We find that setting $\beta=-10^{-6}$ has the additional benefit of prolonging the final inflationary phase. The potential is drawn in Fig.~\ref{fig:pot}. The equations of motion and constraint are
\begin{subequations}
\begin{align}
    a'' + \frac{a}{3} \left( \phi ^{\prime 2} + V \right) =  0 &\;, \label{eoma}\\
    \phi '' + 3 \frac{ a'}{a} \phi' - V_{, \phi} =  0 &\;,\label{eomphi}\\
    a ^{\prime 2}  -1 = \frac{a ^2}{3} \left( \frac{1}{2} \phi ^{\prime 2} - V \right) \;. 
\end{align}
\end{subequations}
A prime denotes a derivative with respect to 
`Euclidean time' $\tau$ (related to the Lorentzian time $\lambda$ through $N\dd t \equiv \dd \lambda \equiv -i \dd\tau$). The classical histories that we are interested in are shown in Fig.~\ref{fig:class}. The bouncing history is obtained from the initial conditions $a(0)=10/\sqrt{5}$, $\phi(0) = -\sqrt{6}$, $\dot\phi(0)=0$. It features an initial expansion phase, followed by a recollapse, a curvature bounce, and finally a second inflationary phase. The scalar field starts on the left side of the potential, midway down, then rolls through the valley at negative values of the potential (which is the reason for the recollapse) and then lands on the plateau on the right hand side of the potential (at positive $\phi$), where inflation can occur. The second history starts at the turning point of the scalar field, near $\lambda=24$, where $\dot\phi=0$. This second history is a purely inflationary one and takes place only on the right-hand side of the potential.

What we would like to assess is whether both of these histories can arise as classical spacetimes from the wave function and, if so, we are also interested in their relative probabilities. This means that we must find the appropriate instanton solutions. Purely inflationary instantons have been studied for a long time and are well known \cite{Lyons:1992ua,Janssen:2020pii}. But instantons featuring a curvature bounce have never been constructed to the best of our knowledge (instantons containing a ghost condensate bounce were found by one of us in \cite{Lehners:2015efa}). 

The strategy for finding such instantons is as follows: we start near the bottom point of the putative instanton solution, known as the South Pole, which we fix to be located at $\tau=0$. 
Thus real values of $\tau$ correspond to Euclidean time. At the South Pole, we have $a(\tau=0)=0$ and $\phi(\tau=0)=\phi_\textrm{SP} \in \mathbb{C}$. Since the South Pole is a regular singular point, we must actually solve the equations of motion numerically starting slightly away from $\tau=0$, using the appropriate series expansions of a solution near $\tau=0$. Then we must see if there exists a point $\tau=\tau_\mathrm{f}$ at which the scale factor and the scalar field simultaneously reach the desired final values $a(\tau_\mathrm{f}) = b \in \mathbb{R}$, $\phi(\tau_\mathrm{f}) = \chi \in \mathbb{R}$. This is an optimisation problem (which is not guaranteed to be solvable), and we must tune both $\phi_\textrm{SP}$ and $\tau_\mathrm{f}$ until the desired final conditions are reached.

\begin{figure}[t]
	\centering
    \includegraphics[width=0.9\textwidth]{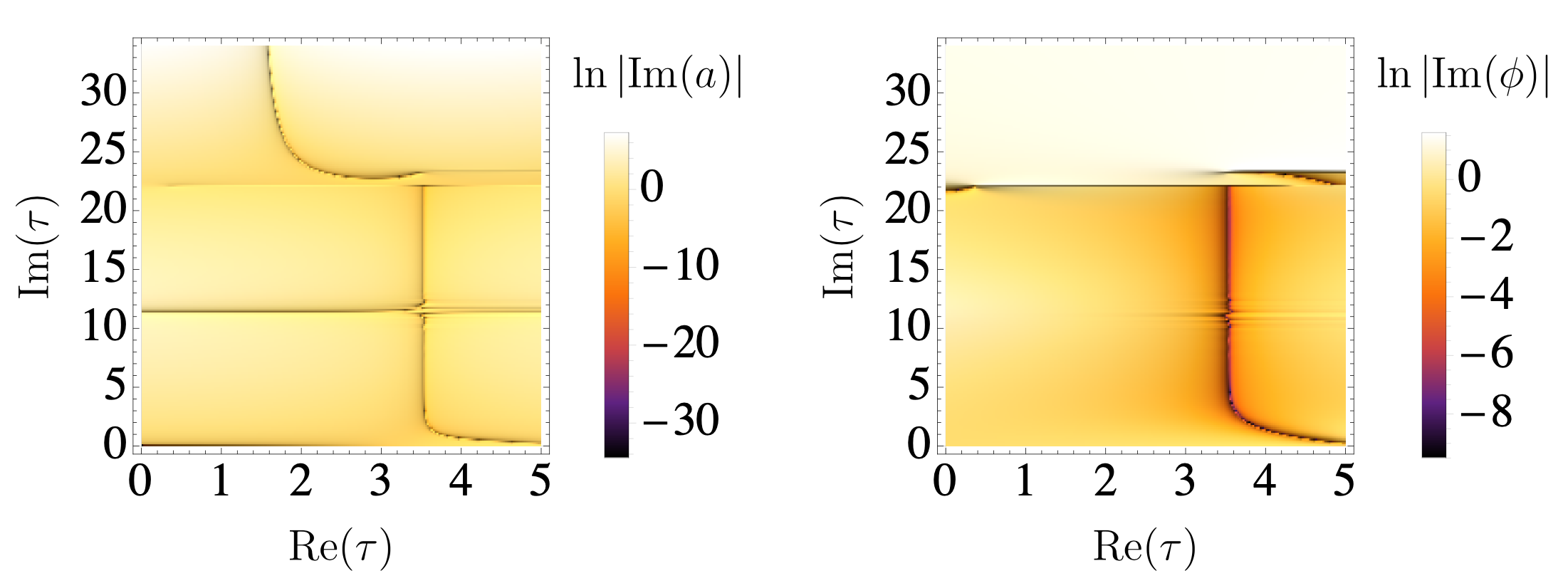}\\
    \includegraphics[width=0.9\textwidth]{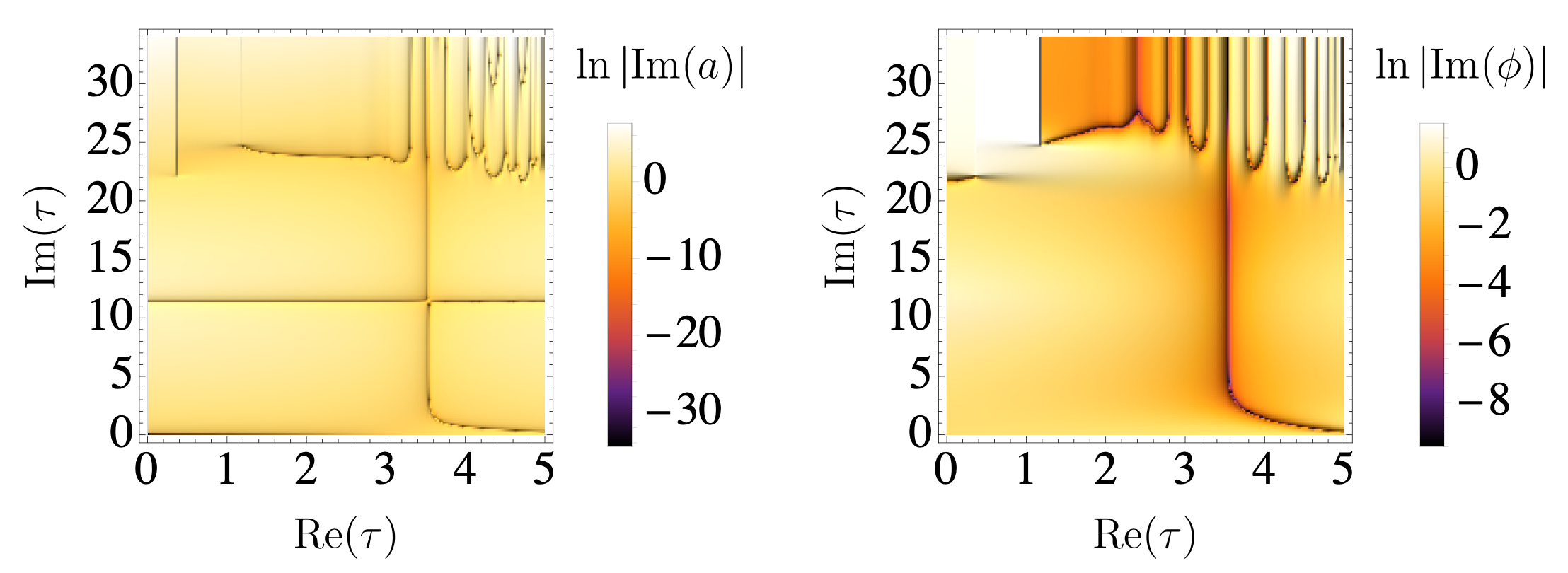}
	\caption{{\footnotesize These graphs show the locus of real field values (dark purple areas) over the complex $\tau$ plane, for the scale factor (left panels) and the scalar field (right panels). The equations of motion are solved starting from the South Pole at $\tau=0$. Top panels: the equations are solved along the imaginary $\tau$ axis first and then over a grid by increasing the real value of $\tau$. Bottom panels: here the equations of motion are solved in the real $\tau$ direction first and then over a grid obtained by increasing the imaginary parts of $\tau$. Wherever the top and bottom panels do not coincide, singularities must be present. This is most evident in the upper regions of the graphs, for $\textrm{Im}(\tau) \gtrapprox 22$. The vertical curves at $\textrm{Re}(\tau) \approx 3.55$ correspond to the approximately classical evolution we are searching for. These graphs were generated with $\phi_\textrm{SP} = -2.878588733 + 0.573260974 i$.}}
	\label{fig:instantongraphs}
\end{figure}

\begin{figure}[t]
	\centering
	\includegraphics[width=0.5\textwidth]{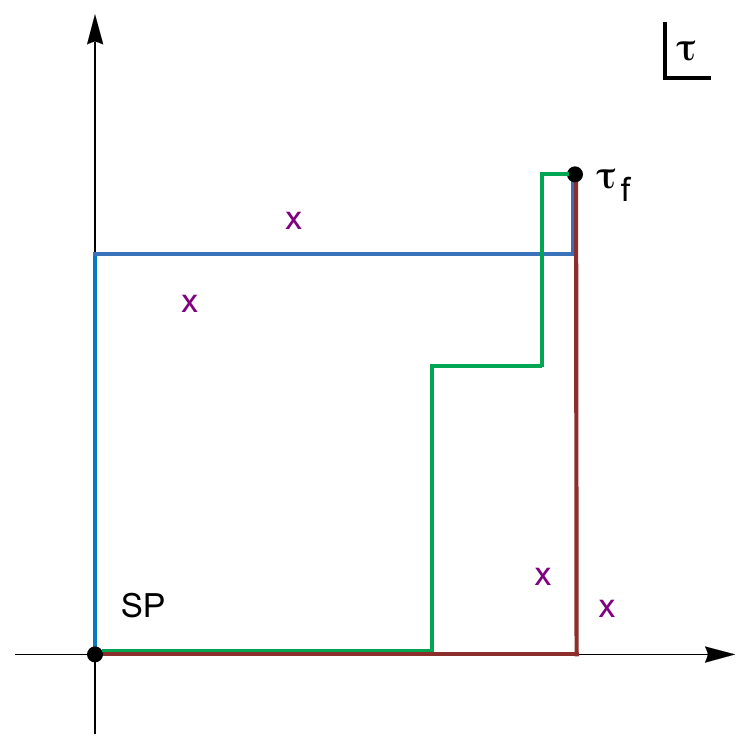}
	\caption{{\footnotesize Contours used for finding and describing instantons: in red the classic `Hawking contour' following the Euclidean time direction first and then the Lorentzian one. In blue a contour that proved more useful when anisotropies were included --- this contour follows the imaginary $\tau$ direction first and then the real $\tau$ direction. The presence of singularities can cause these contours to end up on the wrong sheet of the solution. We found that the best contour for finding bouncing instantons passes between such singularities, and the shape we used is shown in green here.}}
	\label{fig:contour}
\end{figure}

In practice one follows a particular path in the complex $\tau$ plane until at some point the scale factor, say, reaches a real value. Then one uses an optimization algorithm (we used a simple Newtonian algorithm) to progressively tune $\phi_\textrm{SP}$ and $\tau_\mathrm{f}$ until the final conditions are satisfied to the desired level of accuracy. It is in the choice of path that a complication arises. The standard path that used to be considered in early works (see, e.g., \cite{Hawking:1983hj,Lyons:1992ua}) was to follow the Euclidean direction first, along which the solution is approximately described by a (complexified) $4$-sphere until the equator is reached, and then follow the Lorentzian direction until $b, \chi$ are reached. However, when anisotropies are included, it turns out that this path does not work, as there are branch points located near the turning point from Euclidean to Lorentzian contour \cite{Bramberger:2017rbv}. For this reason, it became preferable to use a contour that follows the imaginary $\tau$ axis first and then turns into the Euclidean direction. However, in the present case, we find that additional singularities are present near that turning point too --- see Fig.~\ref{fig:instantongraphs}. Because of both of these concerns, we find that for a contour to work well in general it must pass in between both sets of singularities, although in the present case, because there are no anisotropies present, we could also use the Hawking contour.
The shape of the contour we ended up using is sketched in Fig.~\ref{fig:contour}. We will discuss this contour further in the discussion section, but for now we simply note that with this choice of contour we could find the appropriate `bouncing' instantons.

\begin{figure}[t]
	\centering
    \includegraphics[scale=0.8]{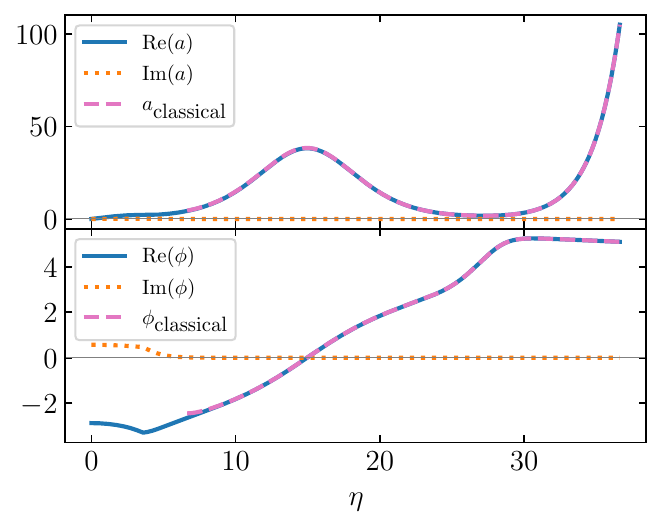}
	\caption{{\footnotesize An example of a bouncing instanton, with field values plotted by following the Euclidean time direction first and then the Lorentzian one, as a function of an affine parameter $\eta$ along the contour. The real parts reach the classical history and then follow it closely. The final values ($b = 105.15254547$ and $\chi = 5.10500692$ obtained at $\lambda=30$) are reached at time $\tau_\mathrm{f} = 3.55358603 + 33.08676411 i$ and with optimised South Pole value $\phi_\textrm{SP} = -2.878588733 + 0.573260974 i$.}}
	\label{fig:instantonbouncing}
\end{figure}

We searched for instantons having final values following the bouncing history shown in Fig.~\ref{fig:class}. Thus we can label the instantons by $\lambda$, the Lorentzian time of the classical history. Note that from the point of view of the wave function, this label has no physical significance, it is just a parameter that we use to specify the final conditions [$b=a(\lambda)$, $\chi = \phi(\lambda)$]. We used integer values of $\lambda$ between $\lambda=0$ and $\lambda=30$. An example of a bouncing instanton, for $\lambda=30$, is given in Fig.~\ref{fig:instantonbouncing}. Here, after the optimization is done, we have plotted the field values following the Hawking contour, i.e., following the Euclidean time direction first and then the Lorentzian one.
(Because there are no anisotropies present here, there are also no singularities between the [green] contour used for the optimisation and the [red] Hawking contour, hence we can safely deform one into the other; the colour refers to the curves in Fig.~\ref{fig:contour}.)
The time parameter along this complex $\tau$ path is called $\eta$ in this case. As one can see, the field values start out complex, but quickly reach real values. Also, the instanton then very closely follows the classical history (dashed pink curve), with the imaginary parts of the fields having become vanishingly small. This is a clear indication that a classical spacetime has been reached. But we will make this statement clearer below.

\begin{figure}[t]
	\centering
    \includegraphics[height=0.27\textheight]{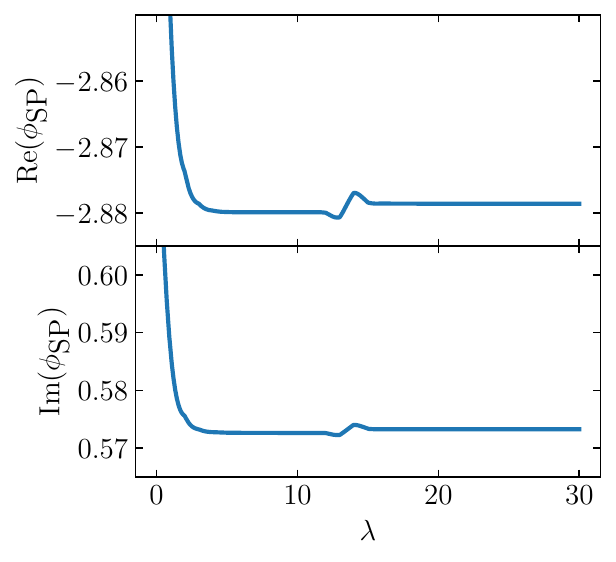}\hspace*{0.03\textwidth}
    \includegraphics[height=0.27\textheight]{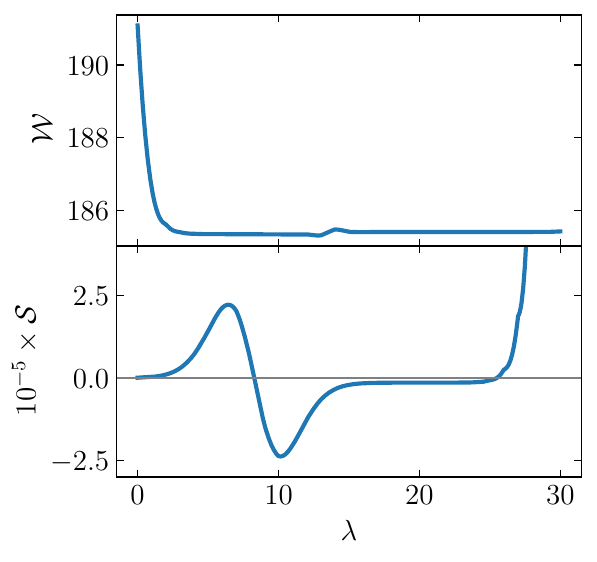}
	\caption{{\footnotesize Optimised South Pole values for bouncing instantons and associated weighting and phase (negative imaginary and real parts of the on-shell action). The curves represent interpolations of our discrete set of numerically optimised instantons at integer values of $\lambda$.}}
	\label{fig:instantonbouncingoptimised}
\end{figure}

For the whole series of bouncing instantons, the optimised South Pole scalar field values are plotted in the left panel of Fig.~\ref{fig:instantonbouncingoptimised}. As one can see, they rather quickly settle to roughly constant values. The only region where there is some change in these values is around $\lambda \approx 13$. By comparing with the classical solution shown in Fig.~\ref{fig:class}, we can see that this corresponds to the recollapse period of the evolution.

The weighting ${\cal W}$ and the phase ${\cal S}$ are shown in the right panel of Fig.~\ref{fig:instantonbouncingoptimised}. The weighting also rather quickly settles to a roughly constant value, so that a definite relative probability for this history can be defined. What is interesting is that the phase is non-monotonic, which is due to the recollapse and bounce parts of the evolution. At late times, when the final inflationary phase is reached, the phase starts growing steeply.

\begin{figure}[t]
	\centering
    \includegraphics[width=0.6\textwidth]{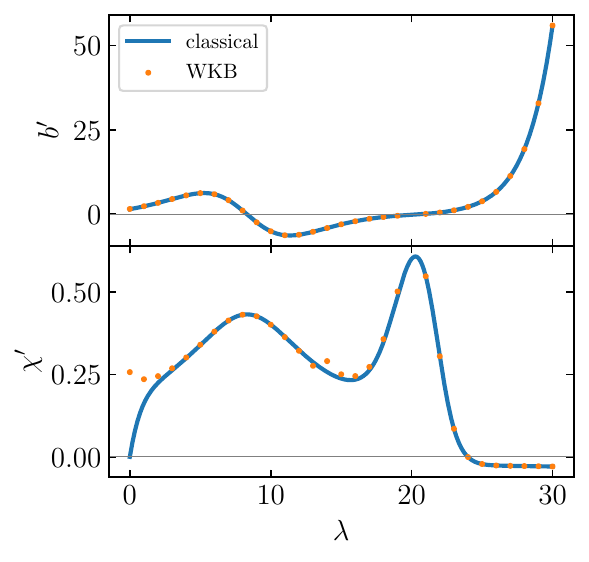}
	\caption{{\footnotesize Field derivatives $b',\, \chi'$ calculated from the wave function (orange dots) compared with the expected classical history (in blue) for bouncing instantons. There is a short period of slight departure from the classical solution during the time of recollapse. In principle this could lead to observable effects.}}
	\label{fig:instantonbouncingmomenta}
\end{figure}

We can now analyse what is perhaps the most interesting aspect of these solutions, namely to what extent they predict a classical spacetime evolution. For this we may compare the momenta calculated from the wave function (using a method of finite differences, which involves constructing instantons with slightly shifted $b$ and $\chi$ values), as specified in Eq.~\eqref{pfromS}, with the momenta evaluated on the classical solution. In fact, we will look directly at the time derivatives of the fields. From the explicit expressions of the momenta, we expect
\begin{equation}
    b'=-\frac{1}{12\pi^2 b}\frac{\partial {\cal S}}{\partial b}\,, \qquad \chi'=\frac{1}{2\pi^2 b^3}\frac{\partial {\cal S}}{\partial \chi}\,.
\end{equation}
In Fig.~\ref{fig:instantonbouncingmomenta} we have plotted the numerically calculated values (orange dots) at each value of $\lambda$ and compared them to the field derivatives calculated from the classical history. As one can see, the two match extremely well as soon as $\lambda >2$, which is an indication that the wave function indeed predicts the occurrence of the classical spacetime evolution that was targeted. There is however a small region around $\lambda = 13$ where the scalar field indicates a small departure from the expected classical evolution. This is during the period of recollapse of the universe and is perhaps not so surprising given that the recollapse period is not a dynamical attractor. However, even during the subsequent bounce, the field evolution is once again very close to the classical solution. Nevertheless, the slight departure from classicality around recollapse shows that in principle one might be able to detect quantum effects in the large scale evolution of the early universe.

\subsection{Competing histories} \label{sec:ch}

\begin{figure}[t]
	\centering
    \includegraphics[width=0.6\textwidth]{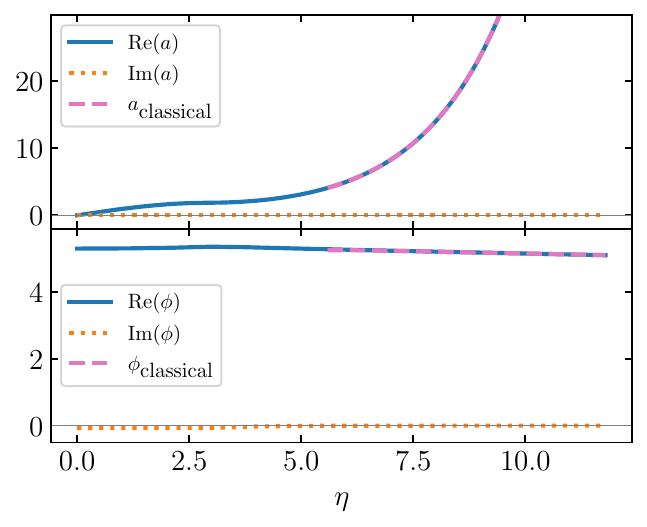}
	\caption{{\footnotesize An example of an inflationary, non-bouncing instanton. The real part reaches the classical history and then follows it closely. The final values ($b = 105.15254547$ and $\chi = 5.10500692$, which are identical to those in Fig.~\ref{fig:instantonbouncing}) are reached at time $\tau_\mathrm{f} = 2.93838843 + 8.85265845 i$ with optimised South Pole value $\phi_\textrm{SP} = 5.30307446 -0.0739360256 i$.}}
	\label{fig:instantonnonbouncing}
\end{figure}

The potential $V(\phi)$ in Eq.~\eqref{eq:pot} allows for a second classical history that foregoes the recollapse and bounce and immediately starts in the inflationary phase on the right-hand plateau at positive $\phi$. This history provides a competition for the bouncing history, as it reaches the exact same field configurations during the inflationary phase. One might thus wonder whether the wave function assigns different probabilities to this history. In order to assess this, we must find the corresponding instantons. These are fairly straightforward to find, as they correspond to small deviations of complexified de Sitter spacetime. In this case, there exist analytic estimates for the scalar field value at the South Pole, namely \cite{Janssen:2020pii} 
\begin{equation}
    \phi_\textrm{SP} \approx \chi - \frac{V_{,\phi}}{V}\frac{\pi}{2} i\,. \label{eq:approx}
\end{equation}
Starting from this value, one can then numerically optimise both $\phi_\textrm{SP}$ and $\tau_\mathrm{f}$ to the desired accuracy. An example of such an inflationary instanton is provided in Fig.~\ref{fig:instantonnonbouncing}. Here the field values are plotted along a Hawking contour, i.e., following the Euclidean time direction first and then the Lorentzian one. One can see that the imaginary parts of the field values decay quickly, and the real parts approach the classical history (in dashed pink) to great accuracy. These instantons lead to a wave function that is of WKB form already after about a single $e$-fold of expansion.

\begin{figure}[t]
	\centering
    \includegraphics[height=0.27\textheight]{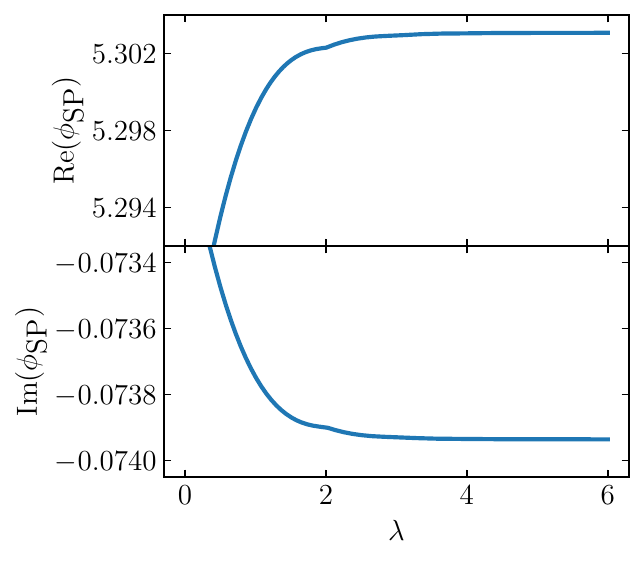}\hspace*{0.03\textwidth}
    \includegraphics[height=0.27\textheight]{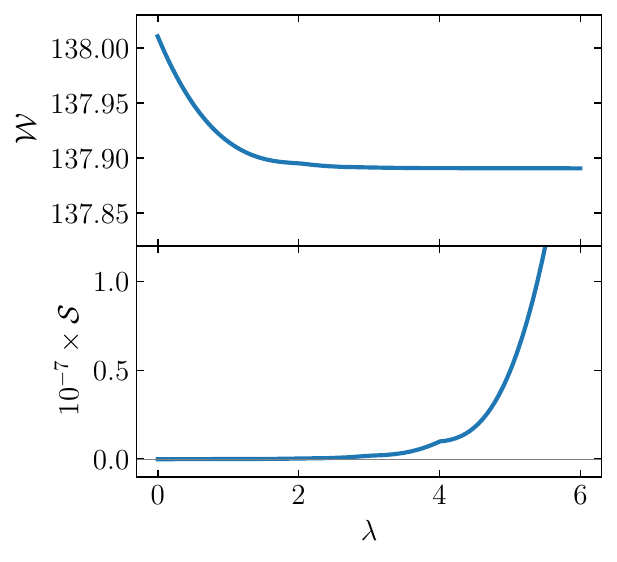}
	\caption{{\footnotesize Optimised South Pole values for `ordinary' (non-bouncing) instantons and associated weighting ${\cal W}$ and phase ${\cal S}$ (respectively negative imaginary and real parts of the complex on-shell action). The curves are interpolations between the numerical solutions as in Fig.~\ref{fig:instantonbouncingoptimised}.}}
	\label{fig:instantonnonbouncingoptimised}
\end{figure}

The optimised South Pole values for the scalar field are shown in the left panel of Fig.~\ref{fig:instantonnonbouncingoptimised} for a series of instantons with increasing values of $\lambda$ and thus also increasing size of the universe. One thing one may note is that the imaginary part of $\phi_\textrm{SP}$ has a different sign compared to the bouncing instantons. This is because the imaginary part is proportional to the derivative of the potential [see Eq.~\eqref{eq:approx}], and thus in some sense the bouncing instantons are anchored on the left side of the potential, while the inflationary, non-bouncing instantons are anchored on the right side, even though their final field evolutions agree with each other. This shows that the instantons remember the origin of the classical history in question, and it is in this manner that they also assign different probabilities to them.

The weighting and phase induced by these instantons are shown in the right panel of Fig.~\ref{fig:instantonnonbouncingoptimised}. One can see that the weighting stabilises very quickly, which is another indication that a classical spacetime has been reached. Meanwhile the phase grows fast as the universe expands during this inflationary phase.

We are now in a position to compare the weightings of the two types of history. The intuition on which the suggestion of Matsui et al.~was based is that the probability of different histories is given in the slow-roll regime by the approximate expression
\begin{equation}
    |\Psi| \approx \exp[\frac{12 \pi^2}{\hbar V(\textrm{Re}(\phi_\textrm{SP}))}]\,,
\end{equation}
i.e., the probability is determined by the potential value that can roughly speaking be thought of as the value reached at the nucleation of the universe. For our two histories, these values differ: the bouncing history starts midway down the left side of the potential at $\phi \approx -2.9$, while the purely inflationary history starts on the right-hand plateau at $\phi \approx 5.3$. This leads to the following expectations for the weightings:
\begin{equation}
    {\cal W}_\textrm{bouncing} \approx \frac{12\pi^2}{V(-2.9)} \approx 199 \,; \quad {\cal W}_\textrm{non-bouncing} \approx \frac{12\pi^2}{V(5.3)} \approx 138\,.
\end{equation}
As expected, the actual value for the inflationary, non-bouncing history is very close to the expected value ${\cal W}_\textrm{non-bouncing} \approx 137.9$ (as can be seen in Fig.~\ref{fig:instantonnonbouncingoptimised}). Meanwhile, the weighting for the bouncing history is not quite as large as expected, but still significantly above the non-bouncing value, ${\cal W}_\textrm{bouncing} \approx 185.4$ (see Fig.~\ref{fig:instantonbouncingoptimised}). Thus the intuition that the bouncing history, which starts lower on the potential, obtains a higher weighting is certainly borne out. Also, because of the smallness of $\hbar$, this implies that the bouncing history in fact obtains an exponentially larger relative probability than the purely inflationary one. As we will discuss in a bit more detail in section \ref{sec:discussion}, the extent to which the bouncing history will remain favoured over the purely inflationary one is thereby not completely settled yet, as the inflationary history is a dynamical attractor all the way, while the recollapse and bounce require initial conditions within a very small basin of field values in order to occur.

The main point to be taken from this analysis is that instantons exist that describe the emergence of a classical spacetime that undergoes a recollapse and a bounce at early times, as a pre-inflationary evolution.

\section{Late bounces} \label{sec:late}

We have just seen that, given a suitable potential, a curvature bounce may have occurred in the early evolution of the universe. We may wonder whether such a bounce could also occur in the future evolution of the universe. After all, if the universe is born in a quantum nucleation event such as the ones described in the previous section, then the average spatial curvature of the universe is positive and could conceivably play an important role once again at future times. What is clear is that this will only be possible if the dark energy, which is currently known to dominate over spatial curvature, decays at some point in the future. Only then could the spatial curvature significantly affect the evolution of (certain regions of) the universe. In the following, we will analyse the required conditions for a recollapse and bounce, give an example and discuss the (lack of) robustness of such bounce solutions.

\subsection{Conditions for collapse and bounce} \label{sec:cb}

Assuming a matter/energy content consisting of a scalar field with a potential (modelling time-dependent dark energy), ordinary pressureless matter with energy density $\rho_\mathrm{m}$, radiation with energy density $\rho_\mathrm{r}$, and also allowing for small shear anisotropy $\sigma^2$ (all treated as perfect fluids), the evolution equations are
\begin{subequations}\label{eq:cosmoEqs}
\begin{align}
    3\left(\frac{\dot a^2}{a^2} + \frac{k}{a^2}-\frac{\sigma_{0}^2}{a^6}\right) & = \frac{1}{2}\dot\phi^2 + V(\phi) + \frac{\rho_{\mathrm{m}0}}{a^3} + \frac{\rho_{\mathrm{r}0}}{a^4}\,, \\
    \frac{\ddot{a}}{a} +2\frac{\sigma_{0}^2}{a^6} & = -\frac{1}{3}\dot\phi^2 + \frac{1}{3}V(\phi) - \frac{1}{6}\frac{\rho_{\mathrm{m}0}}{a^3} -\frac{1}{3} \frac{\rho_{\mathrm{r}0}}{a^4}\,.
\end{align}
\end{subequations}
The subscript $0$ indicates that the quantity is evaluated today when $a=a_0=1$. For convienience, the metric \eqref{eq:ds2k1} is now rewritten in the form $\dd s^2=-\dd t^2+a(t)^2(\dd r^2/(1-kr^2)+r^2\dd\Omega^2_2)$ with $k>0$.

A (re)collapse ($\dot a=0$, $\ddot{a}<0$) can occur when the curvature term competes with all others. As the universe expands, the curvature term naturally grows in importance --- it is only the potential that decreases less fast. Hence, if the potential decays, then the curvature term can come to `dominate.' At the moment of maximal radius $a_\textrm{max}$, one has
\begin{subequations}
\begin{align}
    \frac{3k}{a_\textrm{max}^2} & = \frac{1}{2}\dot\phi^2 + V_\textrm{rec} + \frac{\rho_{\mathrm{m}0}}{a_\textrm{max}^3} + \frac{\rho_{\mathrm{r}0}}{a_\textrm{max}^4} + 3\frac{\sigma_{0}^2}{a_\textrm{max}^6}\,, \\
    V_\textrm{rec} & < \dot\phi^2 +  \frac{1}{2}\frac{\rho_{\mathrm{m}0}}{a_\textrm{max}^3} + \frac{\rho_{\mathrm{r}0}}{a_\textrm{max}^4} +6 \frac{\sigma_{0}^2}{a_\textrm{max}^6}\,.
\end{align}
\end{subequations}
Note that the potential does not need to go negative to induce a recollapse, though this can certainly help precipitate it (see \cite{Andrei:2022rhi} for an exploration of future sudden recollapsing phases). 

When the space collapses, the matter contributions grow fast and can overtake the curvature term. If a bounce is to occur, this must not happen. For this, it is advantageous if the scalar potential grows again, such that it may slow down the rolling of the scalar field and thus diminish its kinetic energy. The analogous conditions at the bounce ($\dot a=0$, $\ddot{a}>0$) are
\begin{subequations}
\begin{align}
    \frac{3k}{a_\textrm{min}^2} & = \frac{1}{2}\dot\phi^2 + V_\mathrm{b} + \frac{\rho_{\mathrm{m}0}}{a_\textrm{min}^3} + \frac{\rho_{\mathrm{r}0}}{a_\textrm{min}^4} + 3\frac{\sigma_{0}^2}{a_\textrm{min}^6}\,, \\
    V_\mathrm{b} & > \dot\phi^2 +  \frac{1}{2}\frac{\rho_{\mathrm{m}0}}{a_\textrm{min}^3} + \frac{\rho_{\mathrm{r}0}}{a_\textrm{min}^4} +6\frac{\sigma_{0}^2}{a_\textrm{min}^6}\,.
\end{align}
\end{subequations}
Note that the growth/decay of the fluid components depends solely on the value of the scale factor. This means that if currently the matter term is more important than the curvature term (which is certainly the case in most regions of the universe, and perhaps everywhere) then the bounce will have to occur at a scale that is larger than the current scale. This statement will be strengthened below.

Let us define
\begin{equation}
    \Omega_{\mathrm{m}0}\equiv\frac{\rho_{\mathrm{m}0}}{3H_0^2}\,,\quad\Omega_{\mathrm{r}0}\equiv\frac{\rho_{\mathrm{r}0}}{3H_0^2}\,, \quad \Omega_{\sigma 0}\equiv\frac{\sigma_0^2}{H_0^2}\,,\quad\Omega_{k0}\equiv-\frac{k}{H_0^2}\,,
\end{equation}
where $H\equiv\dot a/a$ is the Hubble parameter, and let us rewrite the equations of motion \eqref{eq:cosmoEqs} as
\begin{subequations}\label{eq:phiandVasa}
    \begin{align}
        \frac{1}{2H_0^2}\dot\phi(t)^2&=-\frac{1}{H_0^2}\frac{\ddot a(t)}{a(t)}+\frac{1}{H_0^2}\frac{\dot a(t)^2}{a(t)^2}-\frac{\Omega_{k0}}{a(t)^2}-\frac{3\Omega_{\mathrm{m}0}}{2a(t)^3}-\frac{2\Omega_{\mathrm{r}0}}{a(t)^4}-\frac{3\Omega_{\sigma 0}}{a(t)^6}\,,\label{eq:phidot2reconstr}\\
        \frac{1}{H_0^2}V(\phi(t))&=\frac{1}{H_0^2}\frac{\ddot a(t)}{a(t)}+\frac{2}{H_0^2}\frac{\dot a(t)^2}{a(t)^2}-\frac{2\Omega_{k0}}{a(t)^2}-\frac{3\Omega_{\mathrm{m}0}}{2a(t)^3}-\frac{\Omega_{\mathrm{r}0}}{a(t)^4}\,.\label{eq:Vreconstr}
    \end{align}
\end{subequations}
Expressed in this form, we see that given a specification of today's energy densities (the $\Omega$s), the right-hand sides of the equations above depend solely on the scale factor $a(t)$. Thus, given a putative form of $a(t)$, which however must be such that the right-hand side of \eqref{eq:phidot2reconstr} remains positive throughout, one can in principle reconstruct the dependence of the scalar field on time and also reconstruct the associated potential using \eqref{eq:Vreconstr}.

Let us give an example. Consider the functional form
\begin{equation}
    a(t)=A_1\left[\left(1-\frac{A_2}{2A_1H_0}(t-\tilde t_\mathrm{b})\right)\left(1+\frac{1}{C_0+C_1t+C_2t^2}\right)\right]^{-1}\,, \label{eq:adesign}
\end{equation}
with real parameters $A_1,A_2>0$ and $\tilde t_\mathrm{b}>t_0\equiv 1/H_0$. The additional parameters $C_0$, $C_1$, and $C_2$ are fixed in terms of $A_1$, $A_2$, $\tilde t_\mathrm{b}$, and the standard $\Lambda$CDM cosmological parameters today by requiring $a(t_0)=a_0=1$, $\dot a(t_0)=\dot a_0=a_0H_0$, and $\ddot a(t_0)=\ddot a_0$, where using \eqref{eq:phiandVasa} and defining the dark energy equation of state parameter $w\equiv(\dot\phi^2-2V(\phi))/(\dot\phi^2+2V(\phi))$, one can write
\begin{equation}
    \ddot a_0=-\frac{a_0H_0^2}{2}\left((1+3w_0)\left(1-\frac{\Omega_{k0}}{a_0^2}\right)-\frac{3w_0\Omega_{\mathrm{m}0}}{a_0^3}+\frac{(1-3w_0)\Omega_{\mathrm{r}0}}{a_0^4}+\frac{3(1-w_0)\Omega_{\sigma0}}{a_0^6}\right)\,.
\end{equation}
Then, assuming $A_2$ is a small parameter (compared to $H_0^2$) and $\tilde t_\mathrm{b}$ is a large one (compared to $t_0$), one finds
\begin{equation}
    a(\tilde t_\mathrm{b})\approx A_1\,,\qquad \dot a(\tilde t_\mathrm{b})\approx 0\,,\qquad \ddot a(\tilde t_\mathrm{b})\approx A_2\,,\label{eq:atildeb}
\end{equation}
to leading order. Since $A_2>0$, this is approximately a bouncing point, hence we may write $\tilde t_\mathrm{b}\approx t_\mathrm{b}$, $A_1\approx a_\mathrm{b}$, and $A_2\approx \ddot a_\mathrm{b}$.

Let us now use \eqref{eq:phiandVasa} to estimate the scales of a putative future curvature bounce. Clearly, the left-hand side of Eq.~\eqref{eq:phidot2reconstr} is positive. But at the bounce $\dot{a}_\mathrm{b}=0$ and $\ddot{a}_\mathrm{b}>0$, hence the only positive term on the right-hand side is the curvature term $-\Omega_{k0}/a_\mathrm{b}^2$ when $\Omega_{k0}<0$. Radiation and shear are sub-dominant in the universe today, hence we obtain a lower bound on the scale factor at the bounce,
\begin{equation}
    a_\mathrm{b} > \frac{3}{2}\frac{\Omega_{\mathrm{m}0}}{|\Omega_{k0}|} a_0\,. \label{eq:abounce}
\end{equation}
For, e.g., $\Omega_{\mathrm{m}0}=0.3$ and $\Omega_{k0}=-0.01$ (which is roughly the upper bound indicated by current observations depending on the combination probes considered; see, e.g., \cite{Planck:2018vyg,Planck:2018jri,DiValentino:2019qzk,Handley:2019tkm,Vagnozzi:2020dfn,Vagnozzi:2020rcz}) we find that $a_\mathrm{b} > 45 a_0$. With $\Omega_{k0}=-0.001$, it becomes $a_\mathrm{b} > 450 a_0$. Hence a future bounce in our region of the universe could only occur after the universe has first grown by more than about $\mathcal{O}(5)$ $e$-folds at the current rate and then recollapsed favourably. In other words, even under favourable circumstances, a bounce cannot occur earlier than about $70$ billion years from now (this is the justification for $\tilde t_\mathrm{b}\gg t_0$ earlier).
In fixing specific parameters, it is useful to note that positivity of the right-hand side of \eqref{eq:phidot2reconstr} further provides an upper bound on the acceleration at the bounce,
\begin{equation}
    \ddot a_\mathrm{b}<\frac{H_0^2|\Omega_{k0}|}{a_\mathrm{b}}\,.
\end{equation}
Since $a_\mathrm{b}$ must be sufficiently large, $\ddot a_\mathrm{b}$ must be very small (this is the justification for $A_2\ll H_0^2$ earlier).

\begin{figure}
    \centering
    \includegraphics[scale=0.8]{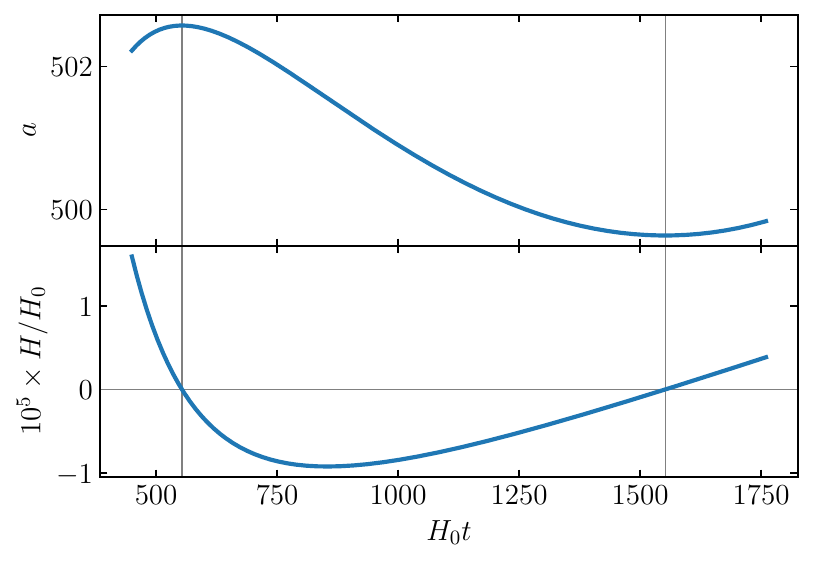}
    \caption{{\footnotesize In these graphs we show the evolution of the scale factor and Hubble rate for the `designed' recollapse and bounce solution given in Eq.~\eqref{eq:adesign}. The specific parameters and initial conditions used are described in the main text.}}
    \label{fig:Hoft_recons}
\end{figure}

\begin{figure}
    \centering
    \includegraphics[scale=0.8]{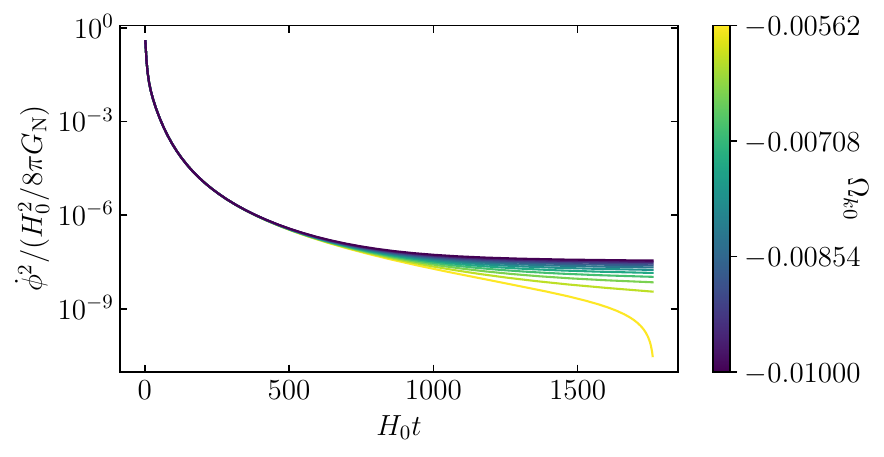}
    \caption{{\footnotesize For a given scale factor ansatz, one may reconstruct the scalar kinetic energy using Eq.~\eqref{eq:phidot2reconstr}. Here we show the reconstruction associated with the scale factor behaviour shown in Fig.~\ref{fig:Hoft_recons}, but for different $\Omega_{k0}$ values. We restricted the values of the spatial curvature such that the kinetic energy remains positive, and thus well-defined, until after the bounce. (For smaller values of the curvature, the kinetic energy becomes negative at some point before the bounce, and the solution is inconsistent.) Note that we explicitly reinserted the units of $8\pi G_\mathrm{N}$.}}
    \label{fig:calK_recons}
\end{figure}

In Fig.~\ref{fig:Hoft_recons}, we plot the scale factor \eqref{eq:adesign} and its corresponding Hubble scale using parameters today that are not too far off from $\Lambda$CDM (specifically, we used $\Omega_{\mathrm{m}0}=0.3$, $\Omega_{k0}=-0.01$, $w_0=-1$, $\Omega_{\mathrm{r}0}=10^{-5}$, and $\Omega_{\sigma 0}=10^{-10}$), and we set $A_1 = 500$, $A_2 = 10^{-5}H_0^2$, and $\tilde t_\mathrm{b} = 1600/H_0$. As one can see in Fig.~\ref{fig:Hoft_recons}, this indeed leads to a recollapse and bounce, as intended, with $a_\mathrm{b}\approx 500$, $\ddot a_\mathrm{b}\approx 10^{-5}H_0^2$, and $t_\mathrm{b}\approx 1600/H_0$, as expected from \eqref{eq:atildeb}. The implied kinetic energy of the scalar field, which is found from \eqref{eq:phidot2reconstr}, is shown in Fig.~\ref{fig:calK_recons}, and we can see that the success of the bounce depends crucially on the amount of spatial curvature (the other parameters are fixed at the same values as in Fig.~\ref{fig:Hoft_recons}). Only for sufficient (positive) spatial curvature does $\dot\phi^2$ remain positive, and only in those cases may the potential be reconstructed.

\begin{figure}
    \centering
    \includegraphics[width=0.99\textwidth]{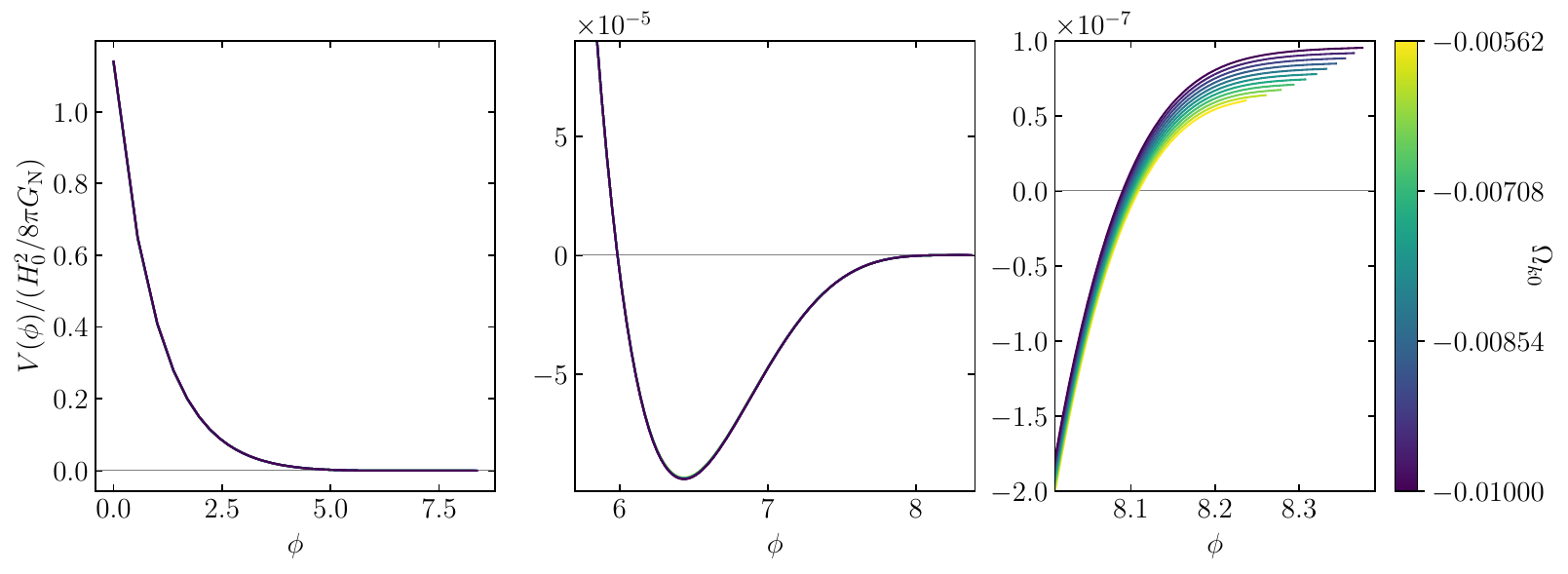}
    \caption{{\footnotesize A reconstruction of the scalar potential with two zoom-ins, given the scale factor evolution of Fig.~\ref{fig:Hoft_recons} and making use of Eq.~\eqref{eq:Vreconstr}. Once again, we explicitely reinserted the units of $8\pi G_\mathrm{N}$. Note that the potential drops steeply to induce a collapse (left panel), but does not need to reach deep negative values (middle panel). The scalar field is then slowed down during recollapse by the rising part of the potential (right panel) and almost comes to rest on the shallow potential plateau at $\phi \sim 8$.}}
    \label{fig:V_recons}
\end{figure}

The reconstruction of the scalar potential is shown in Fig.~\ref{fig:V_recons}, with the middle and right panels providing zoom-ins respectively around the locations where the recollapse and bounce occur. To achieve this, the potential was computed from \eqref{eq:Vreconstr} as a function of time, and $\phi(t)$ was found by integrating the square root of the reconstructed kinetic energy shown in Fig.~\ref{fig:calK_recons}, taking $\phi(2t_0)=0$ as reference scale. The main point to notice in Fig.~\ref{fig:V_recons} is that in order to induce a collapse, the potential must drop steeply. However, there is no need for the potential to reach deep negative values --- in the present example, the potential dips only just below zero. After recollapse, the scalar field runs up the potential and reaches positive, but rather small,  values of the potential. We will see below that this is a general feature in the presence of matter and is distinct from the early universe scenario discussed earlier.

Given that the potential decays during recollapse and grows again during the bounce, we may wonder how high the scalar field could end up after such a bounce. If the inequality \eqref{eq:abounce} is saturated, the Friedmann constraint implies (still neglecting radiation and anisotropies)
\begin{equation}
    \frac{1}{3H_0^2}\left( \frac{1}{2}\dot\phi_\mathrm{b}^2+V(\phi_\mathrm{b}) \right) \approx\frac{|\Omega_{k0}|}{a_\mathrm{b}^2}-\frac{\Omega_{\mathrm{m}0}}{a_\mathrm{b}^3}\approx\frac{4}{27}\frac{|\Omega_{k0}|^3}{\Omega_{\mathrm{m}0}^2}\,,
\end{equation}
which is a very small number. If we now assume that all of the energy density in the scalar field at the bounce point is converted into potential energy shortly after the bounce (i.e., the field rolls up until it is at rest on some sort of plateau), this means the maximal height of that potential is
\begin{equation}
    V_\mathrm{max}\lesssim \frac{4}{9}H_0^2\frac{|\Omega_{k0}|^3}{\Omega_{\mathrm{m}0}^2}\approx(0.0015\,\mathrm{eV})^4\frac{|\Omega_{k0}|^3}{\Omega_{\mathrm{m}0}^2}\,.
\end{equation}
This means that the matter content in the universe prevents the scalar from reaching high potential values after such a bounce. Certainly a future curvature bounce will not bring the scalar field back onto a realistic inflationary plateau.

\subsection{Variation of cosmological parameters under small density perturbations} \label{sec:density}

We have seen that the mere possibility as well as the physical scales of future bounces depend rather sensitively on the relative energy densities of the various constituents of the universe. However, such constituents are not uniformly distributed in the universe, and we may wonder whether bounces could be more likely in other parts of the universe, perhaps those with an under- or overdensity of matter. We analyse this here for small density contrasts, given that the approximation of a Robertson-Walker metric will lose accuracy in the future when large density contrasts are considered.

We generalise the treatment of Goldberg \& Vogeley \cite{Goldberg:2003bw}, who considered voids in a spatially flat background universe. Here we consider a background universe with positive spatial curvature $k$, filled with ordinary matter and dark energy, which for simplicity is modelled as a cosmological constant $\Lambda$. We may assume this simplification because we are only interested in the initial conditions (today) for a potential future bounce. In other words, we are trying to assess whether perturbed regions have conditions favourable for inducing a bounce in the future, which according to the previous subsection mostly amounts to having sufficiently positive spatial curvature today, if dark energy starts decaying only close to today. We can further neglect radiation and shear for our purposes here (see \cite{Bramberger:2019zez} for an exploration of the effects of shear). The background thus satisfies the Friedmann equation
\begin{equation}
    \frac{\dot{a}^2}{a^2}=\frac{\rho_{\mathrm{m}0}}{3a^3}+\frac{\Lambda}{3}-\frac{k}{a^2}\equiv H_0^2\left(\frac{\Omega_{\mathrm{m}0}}{a^3}+\Omega_\Lambda+\frac{\Omega_{k0}}{a^2}\right)\,. \label{eq:FriedBgd}
\end{equation}
As before the $\Omega_{\mathrm{m}0,\Lambda,k0}$ parameters are defined at the present time when $a=a_0=1$, and they sum to~$1$.

We consider the perturbed region to be of spherical extent and as having its own Robertson-Walker metric of the form
\begin{equation}
    \mathrm{d}s^2 = - \mathrm{d}t^2 + a_\delta(t)^2 \left(\frac{\mathrm{d}r^2}{1-k_\delta r^2} + r^2 \mathrm{d}\Omega^2_2 \right)\,,
\end{equation}
where the radial coordinate of the over-/underdense region extends to $0 \leq r \leq r_\delta$. The curvature parameter of the perturbed region is denoted by $k_\delta$ and its spatial curvature radius is then given by $R_\delta = a_\delta/\sqrt{|k_\delta|}$. Note that we do not rescale the coordinates such that $k_\delta \in \{-1,0,1\}$. Rather, we fix the scale by demanding that inhomogeneities are negligible at early times, i.e., $a_\delta \approx a$. The curvature parameter $k_\delta \in \mathbb{R}$ then determines the radius of curvature. The perturbed region satisfies its own Friedmann equation, given by
\begin{equation}
    \frac{\dot{a}_\delta^2}{a_\delta^2}=\frac{\rho_{\mathrm{m} \delta}}{3}+\frac{\Lambda}{3}-\frac{k_\delta}{a_\delta^2}\,,\label{eq:FriedVoid}
\end{equation}
where the perturbed matter density is defined as
\begin{equation}
    \rho_{\mathrm{m} \delta} \equiv \rho_\mathrm{m}(1+\delta) \propto a_\delta^{-3}\,,\label{eq:rhomdelta}
\end{equation}
where the density constrast $\delta(t)$ is negative inside a void and positive inside an overdense region.

At early times, matter dominates both inside and outside of perturbed regions. Then $\delta(t)\propto a(t)$ and consequently $\dot{\delta}/\delta=H$ --- this is a standard result for the growth of perturbations during matter domination \cite{Dodelson:2003ft}, not rederived here --- and by assumption $|\delta |\ll 1$. Taking a derivative of \eqref{eq:rhomdelta} and using $\rho_\mathrm{m}\propto a^{-3}$ and the scaling $\dot\delta=H\delta$ then implies (at early times to leading order in small $|\delta|$)
\begin{equation}
    H_\delta\equiv\frac{\dot{a}_\delta}{a_\delta}=\frac{\dot{a}}{a}\left(1-\frac{\delta}{3}\right)\,,\label{eq:Hubblepert}
\end{equation}
which means that the Hubble rate will be higher in voids and lower in overdense bubbles. Moreover, this implies (again at early times to leading order in small $|\delta|$)
\begin{equation}
    a_\delta=a\left(1-\frac{\delta}{3}\right)\,,
\end{equation}
which we can combine with \eqref{eq:rhomdelta} and the fact that $\rho_{\mathrm{m}\delta}a_\delta^3$ and $\rho_{\mathrm{m}}a^3$ are constants at all times to infer
\begin{equation}
    \rho_{\mathrm{m}\delta}a_\delta^3=\rho_{\mathrm{m}\delta,\mathrm{ini}}a_{\delta,\mathrm{ini}}^3=\rho_{\mathrm{m},\mathrm{ini}}a_\mathrm{ini}^3\big(1+\mathcal{O}(\delta_\mathrm{ini}^2)\big)=\rho_{\mathrm{m}0}\big(1+\mathcal{O}(\delta_\mathrm{ini}^2)\big)\,,
\end{equation}
where the subscript `ini' simply indicates that the quantities are evaluated at some early time when $|\delta|$ is negligibly small. The above thus tells us that
\begin{equation}
    \rho_{\mathrm{m}\delta}=\rho_{\mathrm{m}0}a_\delta^{-3}\quad\implies\quad\frac{1+\delta}{a^3}=\frac{1}{a_\delta^3}\label{eq:rhomdeltarhom0adelta}
\end{equation}
are very good approximations \emph{at all times}.

Subtracting the background Friedmann equation \eqref{eq:FriedBgd} from the perturbed one \eqref{eq:FriedVoid} and using \eqref{eq:Hubblepert} leads to an early-time small-$\delta$ relation between the curvature parameters inside and outside the perturbed region,
\begin{equation}
    k_\delta = k + \frac{5}{3}\frac{\delta}{a}H_0^2 \Omega_{\mathrm{m}0}\,,\label{eq:curvinsideout0}
\end{equation}
where we neglected terms proportional to $\Omega_{\Lambda,k0}\cdot \delta$ as these are clearly sub-dominant. 
We learn three things from this relation. First, since $k_\delta$ is a constant, both sides of \eqref{eq:curvinsideout0} must be constant, with the understanding that $\delta /a$ should be evaluated at some sufficiently early time. Thus, this equation is in fact valid \emph{at all times}. This prompts us to define $\eta\equiv\delta_\mathrm{ini}/a_\mathrm{ini}$, allowing for a slight rewriting
\begin{equation}
    k_\delta = k + \frac{5}{3}\eta H_0^2 \Omega_{\mathrm{m}0}\,. \label{eq:curvinsideout}
\end{equation}
Second, if we would like the perturbed region to have positive spatial curvature ($k_\delta >0$), then we must either have an overdense bubble (in either a flat or positively curved background universe), or we can be in a void with $\eta<0$ as long as there is positive spatial curvature outside ($k>0$) and $|\eta|<3k/(5H_0^2\Omega_{\mathrm{m}0})$. 
Incidentally, this also means that voids in a spatially flat background cosmology always behave like small open universes with $k_\delta<0$. And third, when $\eta<0$ and $k>0$, the radius of curvature inside the void will be larger than outside, or put differently, the curvature energy density will be smaller inside the void than outside. This will make it harder for a void region to bounce, but not impossible. For overdense regions, the reasoning is reversed, and in these regions bounces are more likely to occur.

\begin{figure}
    \centering
    \includegraphics[scale=0.8]{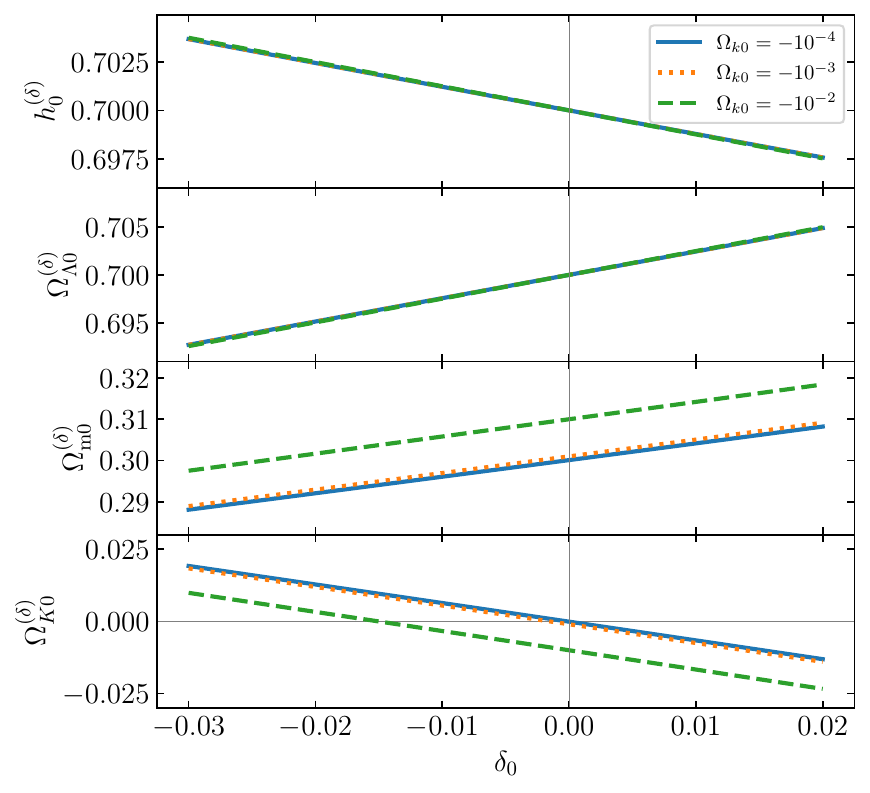}
    \caption{{\footnotesize Under a matter density perturbation $\delta_0$ today, the effective density parameters $\Omega$ change their (local) values. Most importantly, we can see from the lowest panel that even rather small negative density perturbations, leading to void regions, cause those regions to behave like open universes ($\Omega_{K0}^{(\delta)}>0$), even when the background universe has positive spatial curvature, as assumed here (the different levels of background curvature are shown in the inset). In those cases, curvature bounces become impossible. On the contrary, positive density perturbations increase the effective positive curvature and thus lead to more favourable conditions for bounces. Note that the changes in the other density parameters (chosen here to reflect current observations) and in the Hubble rate are rather negligible.}}
    \label{fig:void}
\end{figure}

Figure \ref{fig:void} shows how the various density parameters change as the value of the density contrast today $\delta_0$ is varied, with the assumption that the background universe has positive spatial curvature ($\Omega_{k0}<0$) and using values of the current energy densities that are in agreement with observations. The methodology is the following: starting from standard values for $H_0$, $\Omega_{\mathrm{m}0}$, $\Omega_\Lambda$, and $\Omega_{k0}$ today (to be specific, in Fig.~\ref{fig:void} we set $H_0=70\,\mathrm{km}\,\mathrm{s}^{-1}\,\mathrm{Mpc}^{-1}$, $\Omega_\Lambda=0.7$, $\Omega_{\mathrm{m}0}=1-\Omega_\Lambda-\Omega_{k0}$, and varied $\Omega_{k0}$ over a few orders of magnitude), we integrate the background cosmology \eqref{eq:FriedBgd} backward in time until the scale factor reaches $a_\mathrm{ini}=a_0/10$. Then, starting at that moment in the past, we introduce an initial density perturbation $\delta_\mathrm{ini}$, which is quantified by $\eta=\delta_\mathrm{ini}/a_\mathrm{ini}$, and we solve the perturbed Friedmann equation
\begin{equation}
    3\left(H_\delta^2+\left(k+\frac{5}{9}\eta\rho_{\mathrm{m}0}\right)\frac{1}{a_\delta^2}\right)=\frac{\rho_{\mathrm{m}0}}{a_\delta^3}+\Lambda
\end{equation}
forward until today --- this equation is a good approximation at all times, and it is found by combining \eqref{eq:FriedVoid}, \eqref{eq:rhomdeltarhom0adelta}, and \eqref{eq:curvinsideout}. Spanning different perturbation sizes $\eta$, we get different resulting values for $a_{\delta 0}$ and $H_{\delta 0}$, and in Fig.~\ref{fig:void} we plot the quantities $h_0^{(\delta)}\equiv H_{\delta 0}/(100\,\mathrm{km}\,\mathrm{s}^{-1}\,\mathrm{Mpc}^{-1})$, $\Omega_{\Lambda 0}^{(\delta)}\equiv\Lambda/(3H_{\delta 0}^2)$, $\Omega_{\mathrm{m}0}^{(\delta)}\equiv\rho_{\mathrm{m}0}/(3H_{\delta 0}^2a_{\delta 0}^3)$, and $\Omega_{K0}^{(\delta)}\equiv-(k+5\eta\rho_{\mathrm{m}0}/9)/(H_{\delta 0}^2a_{\delta 0}^2)$ as functions of $\delta_0=(a_0/a_{\delta 0})^3-1$.

The main message to take from Fig.~\ref{fig:void} is that the effective spatial curvature is very strongly affected by the density perturbations, as seen in the lower panel of the figure. This is due to the fact that current observational bounds on the spatial curvature are rather stringent. Then even a small underdensity can cause the corresponding region to evolve like a small open universe, precluding the occurrence of a future bounce. An overdensity, on the contrary, increases the chances for a bounce. Note that the remaining cosmological parameters change much less significantly when such density perturbations are taken into account.

Finally, let us address whether curvature bounces could significantly affect the long-term future of the universe. Bouncing regions may undergo a long subsequent expanding phase, but might be rather empty since they start out pretty empty and then get diluted further. Thus regions that underwent a bounce could potentially lead to long phases of expansion, with very small vacuum energy. If that vacuum energy decays once more, they may recollapse again. Depending on the shape of the scalar potential, they could even bounce repeatedly, though this requires further tuning of the potential. Given that these bounces necessarily occur at very small energy densities, and at very large scales, they cannot reignite the big bang and lead to a realistic cyclic cosmological model.

We should however point out that in a string theoretic setting, the scalar that we were considering could be the dilaton, in which case the coupling constants of nature would also change during a recollapse and bounce phase. It would be interesting to analyse a model of this kind in detail to see how this would impact our analysis of the physical scales associated with curvature bounces. We leave such an analysis for future work.

\section{Discussion} \label{sec:discussion}

The great advantage of the no-boundary wave function is that it can provide us with a potentially complete and non-singular description of the history of the universe. In this paper, we have described a simple setting in which two different histories are present in the wave function and which provide competing alternatives for the very early development of the universe. One such history is a purely inflationary one (presumed to be followed by the standard hot big bang evolution), and the other is an alternative history in which, after nucleation, the universe only expands for a short time before undergoing a recollapse and bounce, only to be followed by an inflationary phase eventually. The potential advantage of this bouncing history is that it obtains an exponentially larger weighting in the no-boundary wave function than the purely inflationary history \cite{Matsui:2019ygj}. 

The model that we analysed contains only currently understood physics and in particular does not rely on violations of the null energy condition to induce a bounce. Rather, the bounce is caused by the positive spatial curvature of the universe, which is a generic feature of no-boundary solutions. The prior recollapse phase is induced by the scalar potential decaying. This is also in line with our current understanding of string theory, where the non-constancy of vacuum energy is an expected feature \cite{Obied:2018sgi}.

A central result of the present paper is the construction of appropriate bouncing no-boundary instantons. Their construction requires some care, as singularities are present in the analytic continuation of the instantons, and the choice of time contour linking the South Pole of the instantons to the final hypersurface becomes non-trivial. What is interesting is that the type of contour that works best in avoiding singularities has a shape that is reminiscent of the contours required by the Kontsevich-Segal allowability criterion, cf.~the examples discussed in \cite{Jonas:2022uqb} (see also \cite{Witten:2021nzp,Lehners:2021mah}). This may be more than a coincidence, and certainly supports the view that allowability is a sensible criterion. 

There are two ways in which the bouncing history could lead to observable consequences. First, the pre-history to inflation implies that the scalar field still has some kinetic energy before the proper inflationary phase starts. This leads to a suppression of fluctuations just at the start of inflation. If inflation lasts only a time marginally longer than required for solving the flatness problem, then one would expect the power on the largest scales in the cosmic microwave background to be suppressed. This is in fact what is seen in WMAP \cite{WMAP:2012nax} and \textit{Planck} \cite{Planck:2018vyg} data (see also \cite{Schwarz:2015cma} and \cite{Piao:2003zm}).

A second, more intrinsically quantum effect is the slight loss of classicality during the recollapse phase. In other words, during recollapse the evolution of the universe shows a slight departure from the solution to the classical equations of motion, and this departure could in principe be observable, although one might have to wait a very long time before the corresponding scales enter our horizon. Still, in principle this exhibits a quantum cosmological effect that is rather distinct.

As for which history, either the purely inflationary one or the bouncing one, is more likely, it is currently hard to give a definite answer. On the one hand, the bouncing history receives an exponentially higher `bare' probability. On the other hand, the bouncing solution requires highly tuned initial conditions, while the purely inflationary solution can be reached from a much larger basin of initial conditions. Only if we knew the measure on the set of all metrics could we ascertain whether the attractor nature of the inflationary history might be able to compensate for the higher weighting of the bouncing history. Unfortunately, we must leave this question for future investigations. We should note that in more elaborate settings, the wave function may well contain a very large number of competing histories, so that this issue will need to be clarified eventually.

Another open question that we have not analysed here is whether the bouncing instantons are themselves allowable in the Kontsevich-Segal sense. And even if they were not, could one construct bouncing instantons that are allowable by having a longer initial expansion phase? This question certainly seems within reach of follow-up work.

In addition to early bounces, we have analysed the conditions under which such curvature bounces might occur in the future evolution of the universe. Here we found that the presence of matter renders the occurrence of bounces much less likely, although not impossible. In fact, we have seen how, given a suitably posited scale factor evolution, one can quite straightforwardly reconstruct the scalar potential that would induce such dynamics. That said, even a small (negative) density perturbation can cause a region of the universe to become effectively open, so that no bounce will be able to take place there. Overdense regions are preferable in that sense, but a question for future work is to what extent the approximation of having a homogeneous and isotropic metric describing that region remains valid as we extrapolate to the rather distant future. Combined with the fact that bounces are required to occur at very low energy scales, we are forced to conclude that curvature bounces are unlikely to significantly affect the future evolution of the universe. 

If curvature bounces played a role in the evolution of the universe, then it appears more likely that they were of importance at early times, where we have the additional benefit that they might have imprinted their signature on the sky.

\section*{Acknowledgements}

JLL gratefully acknowledges the support of the European Research Council (ERC) via the ERC Consolidator Grant CoG 772295 ``Qosmology''.
JQ acknowledges financial support from the University of Waterloo's Faculty of Mathematics William T.~Tutte Postdoctoral Fellowship, and his research at the Perimeter Institute is supported by the Government of Canada through the Department of Innovation, Science and Economic Development and by the Province of Ontario through the Ministry of Colleges and Universities.


\addcontentsline{toc}{section}{References}

\let\oldbibliography\thebibliography
\renewcommand{\thebibliography}[1]{
  \oldbibliography{#1}
  \setlength{\itemsep}{0pt}
  \footnotesize 
}

\bibliographystyle{JHEP2}
\bibliography{references}

\end{document}